\documentclass[12pt]{article}
\usepackage{amssymb}
\usepackage{graphicx}
\usepackage{amsmath}
\newcommand{\be}{\begin{equation}}
\newcommand{\ee}{\end{equation}}
\begin{document}
\baselineskip18pt
\title{Entropy and time: Thermodynamics of diffusion processes}
\author{Piotr Garbaczewski\thanks{
Electronic address: pgar@uni.opole.pl}\\
Institute of Physics,  University  of Opole, 45-052
Opole, Poland}
\maketitle
\begin{abstract}
We  give  meaning to the first and second laws of thermodynamics in
case of mesoscopic  out-of-equilibrium  systems  which are driven by
diffusion-type, specifically Smoluchowski,  processes.  The notion
of  entropy production is analyzed. The role of the Helmholtz
extremum principle is contrasted to that of the more familiar
entropy extremum principles.
\end{abstract}

\vskip0.2cm

\section{Introduction}

 We aim at a consistent thermodynamic description of
diffusion-type  processes which model the dynamics of non-equilibrium systems at the mesoscopic scale,
 \cite{hasegawa,rubi,shizume,munakata,rubi0}.  It is known that given the equilibrium properties of
 a mesoscopic (molecular)  system, it is possible to deduce a stochastic nonequilibrium, albeit near-equilibrium,
 dynamics  in terms of Fokker-Planck equations and their probability  density solutions, \cite{rubi0}.

 We basically go in reverse and abandon any
 prescribed  concept of local or global equilibrium and ask for these thermodynamic properties that
give  account of  a convergence  (if any, this porperty is not
automatically granted) towards an equilibrium state, even if
initially a system is arbitrarily far from equilibrium,
\cite{gar,gar1,gar2}. Our focus is on a quantitative description of
   energy (heat, work, entropy and entropy production) transfer  time rates in the mean, between a
  particle and its thermal  environment.

   We explore the extremum principles which are responsible for the
  large time asymptotic of the process, \cite{glansdorf}.
 Thermodynamic function(al)s, like e.g.
 an internal energy,  Helmholtz free energy and Gibbs-Shannon entropy are inferred,  through suitable averaging,  from
the   time-dependent  continuous probability densities,
\cite{qian,sasa,seifert,rubi0} and \cite{mackey,gar,gar1,gar2}.
Assuming  appropriate (natural) boundary data we
   demonstrate that generically  the corresponding extremum principle amounts to  minimizing the  Helmholtz free
  energy of random motion, see also \cite{hasegawa}.

The following  hierarchy of  thermodynamical systems is adopted: {\it isolated} with no energy
and matter exchange with the environment,  {\it closed} with the energy   but  {\it no } matter exchange and
 {\it open} where energy-matter exchange  is unrestricted.
With the standard text-book wisdom in mind that all isolated systems evolve to the state of equilibrium
 in which  the entropy reaches its maximal value, we focus  our  further
  attention on  {\it  closed} random  systems and their somewhat different   asymptotic features.

A concise resume of a non-equilibrium thermodynamics of {\it closed}
systems comprises the  the basic  conservation laws for   the time
rates of internal energy, heat, work and entropy exchange. The
energy conservation  implies the   $I^{st}$  law  of thermodynamics:
an  internal energy $U$  changes by $dU$ in time $dt$, according to
\begin{equation}
 d{U} = \delta Q + \delta W
 \end{equation}
 where we distinguish the imperfect differentials  by $\delta $. Normally (which will \it
 not \rm
 necessarily be the case in our further discussion) one interprets
 $dU$ as an increase  in internal energy of the system due to
 absorbed heat $\delta Q>0$ and  work  $\delta W
 <0$ performed  by the system  upon its environment.

The $II^{nd}$  law  correlates the   time rates of  entropy, entropy
production and heat exchange  between the system and its
environment:
\begin{equation}
\dot{S}= (\dot{S})_{int} + (\dot{S})_{ext}  \, .
\end{equation}
The    entropy time  rate of  change is here manifestly decomposed
into two contributions:   $(\dot{S})_{int}$   is induced by
irreversible processes that are intrinsic to the system, while
$(\dot{S})_{ext}$  refers to an energy exchange between the system
and its environment.

Since  $(\dot{S})_{int}\geq 0$, this  \it entropy production \rm
term is interpreted as the major signature of  the $II^{nd}$ law,
quite apart form its  specific verbal  formulation.  The remaining
$(\dot{S})_{ext}$ term is related to the  heat exchange via $
(\dot{S})_{ext} dt = \delta Q /T$, where $T$ is the temperature,
\cite{glansdorf,kondepudi}.

We emphasize that neither heat nor work can be interpreted as legitimate thermodynamic functions. Moreover, the very  notion of entropy,
sometimes viewed as a fundamental thermodynamic quantity, appears to be a secondary - derived notion. In the forthcoming statistical description,
this issue will become straightforward, once we shall  relate probabilities and statistics of random events to the (information) entropy notion.

At this point there is no mention of
stationary or steady states, nor any restriction upon the speed of
involved,
 basically irreversible  dynamical process.   For the record, we indicate that
  in case of  a reversible  process we would  have $(\dot{S})_{int} =0$  so that an
   overall  entropy change would arise  solely due to the flow of heat.

Thermodynamical extremum principles are usually invoked in
connection with the large time  behavior of  irreversible processes.
One  looks for direct realizations of the entropy growth paradigm,
undoubtedly valid for  isolated systems, \cite{mackey}, compare e.
g. also a collection of various entropy optimization strategies in
Ref.~\cite{kapur}.

Among a number of admissible  thermodynamic extremum principles, just  for
 reference in the present context,   we single out  a specific one.
If the temperature $T$ and  the available
volume $V$ are kept constant, then
the minimum of the Helmholtz free energy  $F= U - TS$  is
preferred  in the course of the system evolution in time,
 and  there holds  $ \dot{F} = - T(\dot{S})_{int} \leq
0$.

\section{Randomness vs uncertainty:  Boltzmann and  Gibbs-Shannon  entropies}

We know that  a result of an observation of any random phenomenon
cannot be predicted a priori (i.e. before an observation), hence
it is natural to quantify an  uncertainty of this phenomenon. Let
us  consider $\mu =(\mu _1,...,\mu _N)$ as a probability measure
on  $N$  distinct (discrete)    events $A_j, 1\leq j\leq N$
pertaining to  a  model  system. Assume that  $\sum_{j=1}^{N} \mu
_j = 1$ and  $\mu _j =prob(A_j)$  stands for  a probability for an
event $A_j$ to occur in the game of chance with $N$ possible
outcomes.

The expression, whose functional (logarithmic term) provenance may
be traced back to the thermodynamical  notion of  Gibbs entropy,
\begin{equation}
{\cal{S}}(\mu ) = - \sum_{j=1}^N \mu _j \ln \, \mu _j  \,
\label{info}
\end{equation}
stands for  the  measure of the   \it  mean    uncertainty  \rm
  of the possible outcome of the game of chance  and
at the same time quantifies the \it  mean information \rm
 which is  accessible  from an experiment  (i.e.
actually playing the game).

If we identify random   event values   $A_1, ..., A_N$  with labels
for particular
 discrete  "states" of the system, we may interpret  Eq.~(\ref{info}) as a measure of uncertainty
of the  "state"  of the system, \it before \rm   this particular
"state" it is chosen out of the set of  all admissible ones. This
well conforms with the standard  meaning attributed to the Shannon
information  entropy: it is  a measure of the degree of ignorance
concerning which possibility (event $A_j$) may hold true in the set
$\{ A_1, A_2,...,A_N\}$ with  a given a priori probability
distribution $\{ \mu _1, ...,\mu _N\}$.

Notice that:
\begin{equation}
0 \leq  {\cal{S}}(\mu ) \leq  \ln \, N  \, \label{uncertain}
\end{equation}
ranges from certainty (one entry whose  probability equals $1$ and
thus no information  is missing)
  to  maximum uncertainty when a uniform distribution $\mu _j = 1/N$ for all $1\leq  j \leq N$  occurs.
 In the latter situation, all events (or measurement outcomes) are equiprobable and $\log \,N$ sets
maximum for a measure of the "missing  information".

By looking at all intermediate levels of randomness allowed by the
inequalities Eq.~(\ref{info})  we realize that
 the lower is  the Shannon entropy the  less information about "states" of the system   we are missing, i.e. we have
more information  about the system.
 If the Shannon entropy  increases,  we actually loose an information available  about the system.
Consequently, the difference between two uncertainty measures can
be interpreted as an information gain or loss.

Anticipating various thermodynamic connotations (c.f. Boltzmann and
Gibbs entropy  notions)  we must be careful while introducing
(potentially obvious) notions of events, states, microstates and
macrostates of a physical (or biological) system, cf. \cite{gar2}.
The celebrated Boltzmann formula
\begin{equation}
{\cal{S}}=k_B \ln W \doteq  - k_B \ln P  \label{boltz}
 \end{equation}
 sets a link of  entropy  of the  (thermodynamical)  system with the probability $P=1/W$
that an  appropriate "statistical  microstate" can  occur. Here,
$W$ stands for a number of all possible (equiprobable) microstates
that imply  the  prescribed macroscopic (e.g. thermodynamical)
behavior corresponding to a \it fixed \rm value of ${\cal{S}}$.

It is instructive to recall that  if $P$ is  a probability of  an
event i.e. of a particular microstate, then $- \ln P$  (actually,
with  $\log _2$ instead of $\ln $)  may be interpreted
  as "a measure of information produced when one message
  is chosen from the set, all choices being equally likely" ("message" to be identified
  with a  "microstate").  Another interpretation of
  $- \ln P$ is  that of a degree of uncertainty in the trial experiment.

 {\bf Remark 1:} As a pedestrian illustration let us invoke a classic example of a
 molecular gas  in a box which is divided into two halves denoted "$1$" and "$2$". We allow the
 molecules   to be in one of two  elementary  \it  states: \rm
  $A_1$ if a molecule can be found in  "$1$"  half-box and
$A_2$ if it placed in another  half "$2$".  Let us consider a
particular $n$-th  \it macrostate \rm
 of a molecular gas  comprising a  total of $G$ molecules  in a box,  with $n$ molecules
 in the state $A_1$ and $G-n$ molecules in the state $A_2$.
The total number of ways in which  $G$  molecules  can be
distributed between two halves of the box in this prescribed
macrostate,   i.e.  the number $W = W(n)$   of distinct
equiprobable  \it microstates, \rm clearly is $W(n) = G!/[n!
(G-n)!]$. Here,  $P(n) =1/W(n)$ is a probability with which any of
microstates may occur in  a system  bound to "live"   in a given
macrostate. The maximum of $W(n)$  and thus of $k_B\ln W(n) $
corresponds to $N_1=N_2=n$, see e.g. at the "dog-flea" model discussion\cite{dog}.

To get a better insight into the  information-uncertainty
intertwine, let us  consider an ensemble of  finite systems which
are  allowed to appear  in  any of  $N>0$ distinct elementary
states. The meaning of "state" is left unspecified, although an
"alphabet" letter may be  invoked for convenience.

Let us  pick up   randomly  a large sample composed of  $G \gg 1$
single systems, each  one in a certain  (randomly assigned)
state. We  record  frequencies $n_1/G \doteq p_1,...,n_N/G \doteq
p_N$ with which the  elementary  states  of the type $1,...,N $ do
actually occur.
 This sample is a substitute for a "message" or a "statistical microstate" in the previous
 discussion.

Next, we   identify the number of \it  all \rm  possible  samples
of that    fixed  size $G$  which  would   show up    the very
same statistics $p_1,...,p_N$  of elementary states. We interpret
those samples to display  the same   "macroscopic behavior".

  It was the major discovery due to
  Shannon    that  the   number  $W$  of  relevant  "microscopic   states"
   can be approximately  read out from  each single   sample   and  is  directly  related
   to the the  introduced a priori  probability measure $\mu _1,...,\mu _N$, with an identification
    $p_i \doteq \mu _i$ for all   $1\leq i\leq N$,   by the  Shannon
   formula:
 \begin{equation}
  \ln W \simeq   - G  \sum_{i=1}^N  p_i \ln p_i  \doteq  - G \cdot {\cal{S}}(\mu )  \label{shan}
 \end{equation}

On the basis of this formula, we  can consistently introduce
${\cal{S}}(\mu )$ as   the \it   mean information \rm   per each
 ($i$-th)  elementary state of
the $N$-state    system, as encoded in a given  sample  whose
size
  $G\gg 1$ is sufficiently large.

By pursuing the Shannon's communication theory track,
\cite{gar2}, we can  identify states of the model system
with  "messages" (strings)  of  an  arbitrary  length $G >0$
which  are entirely
 composed by  means of the prescribed  $N$  "alphabet" entries (e.g. events or alphabet letters $A_j$
with the previous probability measure $\mu $). Then,
Eq.~(\ref{shan}) may  be interpreted as  a measure of \it
information \rm   per alphabet letter,  obtained \it after \rm  a
particular message (string $\equiv $ state of the model system)
has been received or measured, c.f. our discussion preceding
Eq.~(\ref{shan}).

 In this case, the
 Gibbs-Shannon entropy (by historical reasons we rename Shannon's  ${\cal{S}}(\mu )$  the Gibbs-Shannon) interpolates
 between a maximal information  (one certain event)
and  a minimal information (uniform distribution), cf.
Eq.~(\ref{uncertain}).
 The above discussion may serve as a useful introduction to an issue of the Shannon information workings
 in genomes and DNA sequences, \cite{chen}

Till now, we have considered discrete   probability distributions
and their  uncertainty/delocalization measures  (Gibbs-Shannon
entropy).   The main objective of the present paper is a discussion
of the temporal behavior of Gibbs-Shannon entropy  of a continuous
probability distribution.

We shall focus on continuous probability distributions on $R^+$. The
corresponding Gibbs-Shannon entropy is introduced as follows:
\begin{equation}
\int \rho (s) \, ds = 1  \rightarrow
{\cal{S}}(\rho)=   - \int \rho (s) \ln \rho (s)  dx
\end{equation}

At this point it is instructive to mention that in the realistic (data analysis) framework, one encounters discrete probability
data  that are inferred   from frequency statistics, encoded in various histograms.  Definitely, there are  \it no \rm continuous
probability densities at work. They typically appear  as  computationally  useful continuous approximations  of
discrete probability measures.

The situation becomes involved in case of the corresponding
Gibbs-Shannon entropies, where the approximation issue is delicate.
Even if one follows a pedestrian reasoning,   we can  firmly justify
and keep under control  the limiting behavior, \cite{cover,gar}:
\begin{equation}
  \sum_1^N \mu _j =1 \rightarrow \int \rho dx = 1\, .
\end{equation}
An immediate question is: what can be said about the  mutual  relationship of $ S(\mu ) =  - \sum_1^N  \mu _j  \ln \mu _j $
and $S(\rho)=   - \int \rho (s) \ln \rho (s) ds $ ?

We first observe that  $0 \leq - \sum_1^N \mu _j \ln \mu _j \leq \ln N$ and consider  an interval of length  $L$ on a line  with the a priori chosen
 partition   unit  $\Delta s = L/N$.
  Next, we define: $\mu _j \doteq  p_j \Delta s $  and notice that (formally, we bypass an issue of dimensional quantities)
\begin{equation}
S(\mu ) = - \sum_j  (\Delta s)  p_j \ln p_j  - \ln (\Delta s)
\end{equation}

Let us fix $L$ and allow   $N$ to grow, so that   $\Delta s $  decreases and the partition becomes finer. Then
\begin{equation}
 \ln (\Delta s) \, \,  \leq  \, \,   - \sum_j  (\Delta s)  p_j \ln p_j \, \,
 \leq \, \,  \ln L
\end{equation}
where
\begin{equation}
S(\mu ) + \ln (\Delta s)  =    - \sum_j  (\Delta s)  p_j \ln p_j  \Rightarrow
S(\rho)=   - \int \rho (s) \ln \rho (s) ds
\end{equation}
\vskip0.5cm
$S(\rho )$ is the Shannon  information  entropy  for the probability measure on the
interval $L$.  In the infinite volume $L\rightarrow \infty $ and
infinitesimal grating $\Delta s \rightarrow 0$ limits, the density functional $S(\rho )$  may be
unbounded both from below and above, even non-existent, and seems to  have lost any computationally  useful link with its
 coarse-grained version $S(\mu )$.

 However, the situation is not that bad, if we invoke standard methods \cite{cover,gar} to overcome a dimensional difficulty,
 inherent in the very definition of
$S(\rho )$, while  admitting  dimensional units. Namely, we can from
the start take   a (sufficiently small) partition unit
 $\Delta s$ to have dimensions of length. We allow $s$ to carry length  dimension as well. Then, the dimensionless expression
 for the Shannon entropy of a continuous probability distribution  reads:
 \begin{equation}
S_{\Delta }(\rho)=   - \int \rho (s) \ln [ \Delta s \cdot  \rho (s)] ds
\end{equation}
and all of a sudden, a comparison of   $S(\rho )$ and its
coarse-grained version $S(\mu ))$  appears to make sense. We can
legitimately set estimates for $|S(\mu ) - S_{\Delta }(\rho )|$ and
directly verify the approximation validity of $S(\mu )$ for a
discrete probability  distribution, in terms of the entropy
$S_{\Delta }(\rho )$ for a $\Delta s$-rescaled continuous
probability distribution, when the partition becomes finer.

{\bf Remark 2:} The value  of  ${\cal{S}}(\rho _{\alpha })$ is
$\alpha $-independent  if we consider  $\rho _{\alpha }(x)=\rho (x-
\alpha )$. This reflects the translational invariance of the Shannon
information measure. Let us furthermore investigate an effect of the
scaling transformation. We denote $\rho _{\alpha, \beta }
 = \beta \, \rho  [\beta (x-\alpha )]$, where $\alpha >0, \beta >0$.
  The respective Shannon entropy reads: ${\cal{S}}(\rho _{\alpha ,\beta })  = {\cal{S}}(\rho )  -
 \ln \beta $. An adjustment  $\beta
 \equiv \Delta s$  sets an   obvious  link with our previous discussion.

{\bf Remark 3:}    In the present paper we are interested in
properties of various continuous probability distributions, and \it
not \rm their coarse-grained versions. Therefore our further
discussion will be devoid of any dimensional or partition unit
connotations.  Since negative values of the Shannon entropy are  now
admitted, instead of calling it an information measure, we prefer to
tell about a "probability  localization measure",
 "measure of surprise" or "measure of information deficit".

\section{Helmholtz free energy and its extremum}

Consider an equilibrium state in statistical mechanics, with  $\beta $ as an inverse temperature.
As the i-th microstate we take an energy (level)   $E_i$, $i  \in I$, with a statistical (Boltzmann)   weight $\exp (- \beta E_i)$.
The macrostate is  introduced as follows: choose  a sample    $E \doteq \{ E_{i_1}, E_{i_2},..., E_{i_n},...\}$ and define
the associated
\begin{equation}
F(\beta ) = - {\frac{1}{\beta }} \ln  Z(\beta )
\end{equation}
with a statistical sum (partition function) $Z$
\begin{equation}
 Z(\beta  ) = \sum _E \exp (- \beta E_i) \, .
\end{equation}
An internal energy reads
\begin{equation}
U= - {\frac{\partial }{\partial \beta }} \ln Z(\beta ) = \langle E \rangle \doteq  \sum _i E_i  \exp (- \beta E_i)
\end{equation}
while an entropy notion $S$ with   $T = 1/\beta $ appears   through
\begin{equation}
U - F \doteq T S
\end{equation}

The "maximum entropy principle" may be replaced by  (or in the least-rewritten as)
the "principle of minimum  free energy".
Indeed, let $p_i$ be a probability of occurrence of a microstate $E_i$  in the macrostate configuration $E$, \,  $ \sum  p_i =1$.
 A  { minimum }  of
 \begin{equation}
 F = U - {\beta }^{-1}S =  F [p] = \sum_i ( p_i E_i + {\frac{1}{\beta }} p_i \ln p_i )
\end{equation}
is achieved  for a canonical distribution:
\begin{equation}
p_i = {\frac{1}{Z}} \exp (- \beta E_i) \, .
\end{equation}

Define $ S[p] = -\sum  p_i \ln p_i $  and  $U =  \sum E_i p_i $.
In order to get an equilibrium distribution associated with the Shannon-Boltzman-Gibbs entropy $S$, we need to { extremize} the functional:
\begin{equation}
\Phi [p] = - \sum p_i \ln p_i - \alpha \sum p_i - \beta \sum E_i p_i
\end{equation}
where $\alpha $ and $\beta $ are the Lagrange multipliers.
We have ($p_i^*$ denotes an equilibrium probability, e.g.  an ultimate solution)
\begin{equation}
\delta \Phi [p]  =  0 = [ -\ln p^*_i - 1 - \alpha  - \beta E_i ] \delta p_i
\end{equation}
(with arbitrary variations $\delta p_i$).
Multiply  the result    by $p_i$, sum up, use the constraints (normalization and the fixed internal energy   value) $\rightarrow $
\begin{equation}
\alpha + 1 = S_*  - \beta U_*
\end{equation}
$$\Downarrow
$$
$$
p_i^*  =  \exp [- S_* + \beta  U_*]  \exp (- \beta E_i) =  \exp  \beta (F_*  - E_i)  \doteq  {\frac{1}{Z}} \exp (- \beta E_i)\, .
$$
Notice that we deal here with a discrete probability measure, i.e. the set of $p_i^*$'s such that $\sum p_i^* = 1$.

 $S_*$ is the  Shannon entropy  of this discrete probability  measure.
In view of $F= U - \beta ^{-1}S$,    the Shannon entropy actually     has  been {  maximized} under  the normalization (probability measure)
 and fixed internal energy constraints.
 To be sure that  the above  $F^*$ is indeed  a minimum, let us consider the {  relative Kullback-Leibler} entropy:
\begin{equation}
 K(p,q) \doteq \sum p_i \ln ({\frac{p_i}{q_i}})
 \end{equation}
and use the  measure $ p_* \equiv \{p_i^*\}$   as the reference one (e g. $ q$).
We  have  ( K is a convex   function with a minimum at $0$):
\begin{equation}
K(p,p_*)  =
- S - \sum p_i [ - S_* + \beta U_* - \beta E_i]  =\beta (F- F_*)  \geq 0
\end{equation}
as anticipated before.

In case of discrete probability distributions, in view of Eq.~(16),
a minimum of $F$ is achieved in conjunction  with  a maximum for
$S$. In below, we shall demostrate that such property  is not a
generic feature when continuous probability distributions come into
consideration.

\section{Thermodynamics of  random phase-space motion}

 Now we pass to a detailed  investigation of time-dependent continuous probability distributions and the large
  time behavior of their entropies.
Let us begin  from  a concise resum\'{e} of the  (non-equilibrium)
thermodynamics of \it  closed  \rm  but non-isolated systems. The
laws of thermodynamics may be reproduced in the
form\cite{kondepudi}: $ dU = \delta Q + \delta W$ and $dS= d_{int}S
+ d_{ext}S$,  where $d_{int}S\geq 0$ and  $d_{ext} S = \delta Q /T$.

With respect to the large time behavior, the following extremum principles for  irreversible processes are
typically invoked:\\
(1)  $U$ and $V$ (volume) constant  $\rightarrow $  maximum of
entropy is preferred: $d_{int}S =   TdS - \delta Q \geq 0$,
 together with  a minimum for the entropy production:
${\frac{d}{dt}} \left({\frac{d_{int}S}{dt}}\right) <0$\\
(2)  $S$ and $V$ constant $\rightarrow $ { minimum internal  energy} is preferred:
$dU = - Td_{int}S \leq 0$.\\
(3)  $T$ and $V$ constant $\rightarrow $ { minimum  of}  $F= U - TS$ (Helmholtz free energy)
 is preferred: $ dF = - Td_{int}S \leq 0$. \\
(4) Further principles  refer to the minimum of the Gibbs free  energy and this of enthalpy (we skip them).\\
The Helmholtz extremum principle will be of utmost importance in our
further discussion, as opposed to more  traditional  min/max entropy
principles.

 We are interested not only in the existence of an
extremal  probability density,  but also  at an approach  of $\rho
(x,t)$  towards such a stationary  density in the course of time.
Then the varied time dependent properties of the Helmholtz free
energy,  Gibbs-Shannon and Kullback-Leibler entropies will be of
interest.

Let us  consider a  phase-space diffusion process governed
by the Langevin equation:
\begin{equation}
m\ddot{x}  + m\gamma \dot{x} = - \nabla V(x,t) + \xi(t) \label{motor}
\end{equation}
with standard assumptions about properties of the white noise:
$\langle \xi (t)\rangle =0, \,   \langle \xi (t)\xi (t')\rangle =
\sqrt{2m\gamma k_BT}\,  \delta(t-t')$.
Accordingly, the pertinent phase-space density  $f=f(x,u,t)$ is a
solution of the Fokker-Planck-Kramers equation with suitable
initial data:
\begin{equation}
{\frac{\partial }{\partial t}}f(x,u,t) =
\end{equation}
$$
\left[ - {\frac{\partial }{\partial x}} u + {\frac{\partial
}{\partial u}} \left(\gamma u + {\frac{1}m} \nabla V(x,t)\right) +
{\frac{\gamma k_BT}m} {\frac{\partial ^2}{\partial u^2}}\right] f
$$
Let us define  the Gibbs-Shannon entropy $ {\cal{S}} = {\cal{S}}(t)$
of a continuous probability distribution :
$$ {\cal{S}}(t) = - \int dx\, du f \ln f  = - \langle \ln f \rangle
$$
(By dimensional reasons we should insert a factor $h$ with physical
dimensions of the action under the logarithm, i.e. use $\ln (h f)$
instead of $\ln f$, but since we shall  ultimately  work with  time
derivatives, this step  may be safely
skipped.)

An internal energy  $U$ of the stochastic process reads
$$E(x,u,t) = {\frac{mu^2}2} + V(x,t) \rightarrow  U= \langle E\rangle  $$
and the $I^{st}$ law   takes the form
\begin{equation}
T (\dot{S})_{ext}  +  \langle
\partial _t V\rangle  = \dot{U}
\end{equation}
 where $\langle
\partial _t V\rangle$, if positive,  is interpreted as the time rate of  work externally performed \it  upon \rm the
system. If negative, then we would deal with work performed \it by
\rm  the system.

Furthermore, let us introduce an obvious analog  of the Helmholtz
free energy:
$$F\doteq \langle E + k_BT \ln f\rangle =   U-TS$$
so that
\begin{equation}
\dot{F} - \langle
\partial _t V\rangle =  T (\dot{S})_{ext} - T\dot{S}     = -
T (\dot{S})_{int} \leq 0 \, .
\end{equation}
The above result  is a direct consequence of the Kramers equation.
Under suitable assumptions concerning the proper behavior of
$f(x,u,t)$ at $x,u$ integration boundaries (sufficiently rapid decay  at
infinities) we have \cite{shizume}
$$
T (\dot{\cal{S}})_{ext} = \gamma (k_BT - \langle mu^2 \rangle )
$$
$$\dot{\cal{S}} = \gamma \left[ {\frac{k_BT}m} \langle
\left({\frac{\partial \ln f}{\partial u}} \right)^2 \rangle  - 1\right]
$$
 and thence,  the $II^{nd}$ law

\begin{equation}
 - {\frac{\gamma }m} \langle
\left(k_BT{\frac{\partial \ln f}{\partial u}} + mu\right)^2\rangle =
-T (\dot{S})_{int} \leq 0   \, .
\end{equation}
 As  a  byproduct of the discussion  we have $\dot{F} \leq \langle
\partial _t V\rangle $.
For time-independent $V=V(x)$ we deal with  the  standard
$F$-theorem (the
 extremum principle pertains to  the Helmholtz free energy $F$  which is minimized in the course of
  random motion.

The above discussion encompasses both the forced and   unforced (free) Brownian motion. When $V(x)\equiv 0$, then
 no asymptotic  state of  equilibrium (represented by a probability density) is accessible, the motion is sweeping.
In the forced case, we assume a priori an  existence of a unique
stationary state, c.f. \cite{mackey,mackey1}, for  the  above
phase-space
 random dynamics:
$$
f_{*}(x,u) = {\frac{1}Z} \exp \left[ - {\frac{E(x,u)}{k_BT}}\right] \, .
$$
In this case,  the time rate of the  conditional Kullback-Leibler entropy:
\begin{equation}
{\cal{H}}_c(f_t|f_{*}) = - \int f \ln {\frac{f}{f_{*}}} dx du =
\end{equation}
$$
{\cal{S}}(t) - \ln Z - {\frac{\langle E(x,u)\rangle }{k_BT}}
$$
directly  appears in the  $F$-theorem:
\begin{equation}
k_BT \dot{\cal{H}}_c = - \dot{F}=   +T (\dot{S})_{int}\geq 0
\end{equation}
The (negative definite)  conditional entropy  grows monotonically
towards its maximum at $0$. Notice  that  $(\dot{S})_{int} \geq 0$,
but neither $\langle
\partial _t V\rangle  $ nor $\dot{S}$ need to be positive  and may show
quite complicated patterns of temporal behavior,
\cite{mackey,mackey1}
 and \cite{gar,gar1}. (Both $f_{*} $  and ${\cal{H}}_c$ are
 non-existent in case of  free Brownian motion.

Let us point out that the above discussion is sufficiently general to include a number of currently fashionable
problems, like e.g. that of  molecular motors. To see an obvious link it suffices to mention a  typical "Brownian motor input"
i.e. an explicit functional  form of the  time-dependent driving  component of the exerted  force and its  conservative term
 in Eq.~(\ref{motor}), c.f. \cite{luczka}. As an example we may consider:
 \begin{equation}
m\ddot{x}  + m\gamma \dot{x} = - \nabla V(x,t)  + a \cos (\Omega t)  + F + \xi(t)
 \end{equation}
 where $F$ is a constant external force,  and the spatially  periodic rachet (broken reflection symmetry) potential $V(x)$ is adopted.
An example of the ratchet potential is: $V(x)= V_0[sin(2\pi x) + c_1 \sin(4\pi x) + c_2 \sin (6\pi x)]$.

\section{Thermodynamics of the Smoluchowski process}

Analogous  thermodynamical features are encountered in   spatial random motions, like  e.g. standard
 Smoluchowski processes and their generalizations.
Let us consider
\begin{equation}
\dot{x} = b(x,t) +A(t)
\end{equation}
with $\langle A(s)\rangle =0 \, , \,  \langle A(s)A(s')\rangle
= \sqrt{2D} \delta (s-s')$.

Given an initial   probability density $\rho_0(x)$. We know that the diffusion process drives this density in accordance with
 the  Fokker-Planck equation
\begin{equation}
\partial _t\rho = D\triangle \rho -  \nabla \cdot ( b \rho ) \, .
\end{equation}
We introduce  $u = D\ln \rho $ and $v=b - u$ which obeys $\partial _t \rho = - \nabla (\rho v)$.\\
The Gibbs-Shannon  entropy   of $\rho $
\begin{equation}
{\cal{S}}(t)  = -\langle \ln \rho \rangle
\end{equation}
  typically is not a conserved quantity. We impose  boundary restrictions
that $\rho, v\rho, b\rho $ vanish  at spatial infinities or  other  integration interval borders.
We consider:
\begin{equation}
  D \dot{\cal{S}}  =  \left< {v}^2\right>
    -  \left\langle {b}\cdot {v}
 \right\rangle  \label{balance}  \, .
\end{equation}
We may   pass to time-independent drift fields
  and  set  $ b = \frac{f}{m\gamma }$,  $j \doteq v\rho $,  $ f = - \nabla V $  plus   $D=k_BT/m\gamma $. Then:
\begin{equation}
\dot{\cal{S}} = (\dot{\cal{S}})_{int} + (\dot{\cal{S}})_{ext}
\end{equation}
where
\begin{equation}
k_BT  (\dot{\cal{S}})_{int}  \doteq m\gamma \left<{v}^2\right> \geq
0
\end{equation}
stands for the {  entropy production} rate, while
\begin{equation}
k_BT  (\dot{\cal{S}})_{ext}  =  -  \int {f} \cdot {j}\, dx = -
m\gamma \left\langle {b}\cdot {v}
 \right\rangle
\end{equation}
 (as long as negative  which is not a must)  may be  interpreted as the  {  heat dissipation rate}:$ - \int {f}\cdot {j}\,  dx$.

In view  of $j = \rho v = {\frac{\rho }{m\gamma }} [ f - k_BT \nabla \ln \rho ] \doteq  - {\frac{\rho }{m\gamma }}\nabla \Psi
$  i.e.  $v= - (1/m\gamma ) \nabla \Psi $ and  $f=-\nabla V$, we can  introduce\\
\begin{equation}
\Psi = V + k_BT \ln \rho
\end{equation}
whose mean value stands for   the {   Helmholtz free  energy} of   the random  motion
\begin{equation}
F \doteq \left< \Psi \right> = U - T S \, .
\end{equation}
Here  $S \doteq k_B {\cal{S}}$ and an internal energy is $ U =
\left< V\right>$. Since we assume  $\rho $  and $\rho V v$ to vanish
at the integration volume boundaries,  we get
\begin{equation}
\dot{F}  =     - (m\gamma )
 \left<{v}^2\right> = - k_BT (\dot{\cal{S}})_{int} \leq 0 \, . \label{helm}
\end{equation}
Clearly, $F$ decreases as a function of time  towards its  { minimum},
or  remains constant.

Let us consider the stationary regime   $\dot{\cal{S}} =0$
associated with an (  a priori assumed to exist, \cite{mackey})
invariant density $\rho _{*}$.
 Then,    $$b=u = D \nabla  \ln \rho _{*} $$ and
\begin{equation}
 -(1/k_BT)\nabla V = \nabla \ln\, \rho _{*}  \Longrightarrow \rho _{*} = {\frac{1}Z} \exp[ - V/k_BT]\, .
 \end{equation}
Hence
\begin{equation}
\Psi _{*} = V + k_BT \ln \rho _{*}  \Longrightarrow \langle \Psi _{*} \rangle =
 - k_BT \ln Z  \doteq  F_{*}
 \end{equation}
   with   $Z= \int \exp(-V/k_BT) dx$.
 $F_*$ stands for   a  minimum  of  the time-dependent  Helmholtz
free  energy $F$. Because of
\begin{equation}
Z= \exp (-F_*/k_BT)
\end{equation}
 we have
\begin{equation}
\rho _* =
\exp[(F_* - V)/k_BT]
\end{equation}

Therefore,   the {  conditional  Kullback-Leibler   entropy},
of the density $\rho $    relative to an equilibrium density $\rho _*
$ acquires the form
\begin{equation}
 k_BT {\cal{H}}_c  \doteq  - k_BT \int \rho \ln
({\frac{\rho }{\rho _*}})dx = F_* - F \, .
\end{equation}

In view of the concavity property of
the function $f(w) = - w\ln w$,  ${\cal{H}}_c$ takes only
negative values, with  a maximum at $0$. We have   $F_*\leq F$ and
 $k_BT \dot{\cal{H}}_c = - \dot{F}  \geq 0$.  ${\cal{H}}_c$  is bound to  grow monotonically
 towards $0$,  while  $F$ drops down to   its
 minimum  $F_*$  which is reached  for $\rho _*$. The Helmholtz free
 energy minimum, in the present context (and  in contrast  to the
 previously described case of discrete probability measures),
 remains divorced from any extremal property of the Gibbs-Shannon
 entropy. Only the Kullback-Leibler  entropy shows up an expected asymptotic  behavior. See e.g. also \cite{mackey,mackey1}.

\section{Outlook}

 Standard   (thermodynamical) notions of   entropy are  basically  introduced  under  equilibrium
 conditions and are  not  considered  in the time domain. Our discussion was tailored specifically to  non-equilibrium
 systems and processes.
 Any conceivable idea of approaching the state of equilibrium,  or passing from one such state to another (steady) state,
always  involves the time dependence and the related, often rapid, non-equilibrium
dynamical process.

The major tool invoked in connection with  both  equilibrium and
non-equilibrium phenomena is that of  Gibbs-Shannon entropy whose
definition directly  involves time-dependent probability
distributions. However, let us recall that  except for the
thermodynamical  Clausius case, the very notion of entropy is
non-universal and purpose-dependent, \cite{gar1}. Our entropy choice
has served a  concrete purpose: encompassing  a temporal
  behavior of specific  probability distributions associated with diffusion-type processes.

The  sole entropy methods are neither exclusive nor sufficient to
give full account of the asymptotic properties of diffusion
processes. Additional inputs pertaining the regularity properties of
solutions of Fokker-Planck equations are necessary to guarantee an
existence of  a stationary solution and  to demonstrate that any
other solution of the  pertinent equation must  finally decay to the
stationary one in the large time asymptotic.

For standard diffusion-type processes, we have discussed, the
standard  min/max entropy principles do not literally work,
\cite{kapur}. It is the Helmholtz free energy which shares   proper
extremal behavior.
   On the other hand it is the  conditional  Kullback-Leibler entropy which  (together with its time rate)   stays in close affinity  with the
   Helmholtz free energy of the diffusion process  and with the involved  entropy production.

  The advantage of our methodology is an explicit insight into the
  temporal  behavior of various thermodynamics functionals whose definition is
  normally restricted to equilibrium(or near-equilibrium)  phenomena. The conceptual
  meaning of the  Helmholtz  free energy, or Gibbs-Shannon entropy
  is consistently elevated to the time-domain, for far from equilibrium systems. The auxiliary  notions of work and heat transfer rates
  have received a  transparent  interpretation as well.

\end{document}